\pdfoutput=1
\documentclass[]{smule}
\usepackage[round,authoryear]{natbib}
\usepackage{hyperref}
\usepackage[colorinlistoftodos]{todonotes}
\usepackage{colortbl}
\usepackage{nicematrix}
\usepackage{amssymb}
\usepackage{amsmath}
\usepackage{booktabs}
\usepackage{adjustbox}
\usepackage{soul}
\usepackage[inline]{enumitem}
\usepackage{booktabs}
\usepackage{color}
\usepackage{xcolor}
\usepackage{bbding}
\usepackage{listings}
\usepackage{tikz}
\usepackage{multirow}
\usepackage{pgfplots}
\usepackage{graphicx}
\usepackage{array}
\usepackage{rotating}
\usepackage{multicol}
\usepackage{xspace}
\usepackage{tablefootnote}
\usepackage{libertinus}

\makeatletter
\AtBeginDocument{

}
\makeatother

\definecolor{lightblue}{rgb}{0.68, 0.85, 0.9}

\def\methodnamelong{PromptReverb} 
\def\methodnameshort{PromptReverb} 

\title{PromptReverb: Multimodal Room Impulse Response Generation Through Latent Rectified Flow Matching}

\author[3\dagger\ddagger]{Ali Vosoughi}
\author[1\ddagger]{Yongyi Zang}
\author[2\dagger]{Qihui Yang}
\author[4\dagger]{Nathan Paek}
\author[1]{Randal Leistikow}
\author[3]{Chenliang Xu}

\affiliation[1]{Smule Labs}
\affiliation[2]{University of California, San Diego}
\affiliation[3]{University of Rochester}
\affiliation[4]{Stanford University}

\contribution[\dagger]{Work done during internship at Smule}
\contribution[\ddagger]{These authors contributed equally.}

\abstract{
Room impulse response (RIR) generation remains a critical challenge for creating immersive virtual acoustic environments. Current methods suffer from two fundamental limitations: the scarcity of full-band RIR datasets and the inability of existing models to generate acoustically accurate responses from diverse input modalities. We present \textit{\methodnamelong}, a two-stage generative framework that addresses these challenges. Our approach combines a variational autoencoder that upsamples band-limited RIRs to full-band quality (48 kHz), and a conditional diffusion transformer model based on rectified flow matching that generates RIRs from descriptions in natural language. Empirical evaluation demonstrates that \textit{\methodnamelong} produces RIRs with superior perceptual quality and acoustic accuracy compared to existing methods, achieving 8.8\% mean RT60 error compared to $-$37\% for widely used baselines and yielding more realistic room‑acoustic parameters. Our method enables practical applications in virtual reality, architectural acoustics, and audio production where flexible, high-quality RIR synthesis is essential.
}

\begin{document}

\maketitle

\section{Introduction}
\label{sec:intro}

The pursuit of seamless virtual experiences across VR, gaming, and remote collaboration increasingly demands perceptually convincing environments that extend beyond visual fidelity to encompass spatial audio, which establishes presence through room impulse responses (RIRs) that encode sound propagation from source to listener \citep{liang2024language}. Despite RIRs' crucial role in enabling realistic audio synthesis through convolution with dry signals, their generation remains fundamentally constrained by the trade-off between computational tractability and perceptual quality.

Traditional physics-based approaches, including ray tracing and image-source models \citep{allen1979image,krokstad1968ray,zhao2025ta,liu2025hearing,lyu2025temporal}, provide physically accurate simulation but require detailed geometric and material specifications while remaining computationally intractable at high frequencies necessary for real-time applications. This computational burden has motivated a shift toward learning-based methods that bypass explicit physics simulation. Image2Reverb \citep{singh2021image2reverb} pioneered single-image RIR synthesis but remains critically dependent on depth estimation accuracy, while recent multi-modal approaches such as AV-RIR \citep{ratnarajah2024avrir} and MEAN-RIR \citep{meanrir2025} achieve superior accuracy through panoramic imagery fusion, yet require specialized 360° capture equipment impractical for consumer deployment. Parametric alternatives like FAST-RIR \citep{ratnarajah2021fast_rir} and geometry-dependent MESH2IR \citep{anton2022mesh2ir} offer computational efficiency but demand either precise acoustic parameter specification or detailed 3D meshes, respectively, creating significant barriers to accessibility.

Beyond computational and input constraints, existing approaches face two critical limitations preventing practical deployment. First, the scarcity of high-quality, full-band RIR datasets severely constrains model training; most datasets contain either band-limited recordings (often below 24 kHz) with paired multimodal descriptions, or synthetic RIRs from physics simulation that fail to capture real acoustic complexity. This directly impacts perceptual quality, particularly in high-frequency content crucial for spatial localization and timbral accuracy. Second, despite recent diffusion-based methods for RIR completion or interpolation \citep{delatorre2025diffusionrir,lin2025decor}, no existing approach supports natural language conditioning for complete RIR generation, which is crucial for creative use. Current text-based systems either require technical parameter specification \citep{arellano2025room} or operate as modification tools within manually selected parametric constraints \citep{richter2025score}, rather than enabling intuitive, free-form textual control over acoustic properties.

In this work, we present \textit{\methodnamelong}, a two-stage generative framework that addresses these fundamental limitations through architectural decoupling: (1) a variational autoencoder (VAE) that performs neural upsampling from band-limited to full-band RIRs (48 kHz), enabling us to leverage existing band-limited datasets while producing perceptually complete outputs, and (2) a conditional diffusion transformer based on rectified flow matching \citep{liu2022rectifiedflow} with transformer architectures \citep{peebles2023dit} that generates band-limited RIRs from natural language descriptions. To create diverse textual training data without manual annotation, we developed a caption-then-rewrite pipeline leveraging vision language models for initial scene description followed by LLM-based creative rewriting.

Through empirical evaluation, we demonstrate that our system achieves 8.8\% mean RT60 error compared to 37\% for existing baselines when conditioned on natural language descriptions, representing the first method to synthesize complete RIRs from free-form textual input while eliminating requirements for panoramic captures, depth estimation, 3D geometry, or acoustic parameter specification.\footnote{Audio examples are available at \url{https://ali-vosoughi.github.io/PromptReverb/}.}

\begin{figure*}
    \centering
    \includegraphics[width=0.95\linewidth]{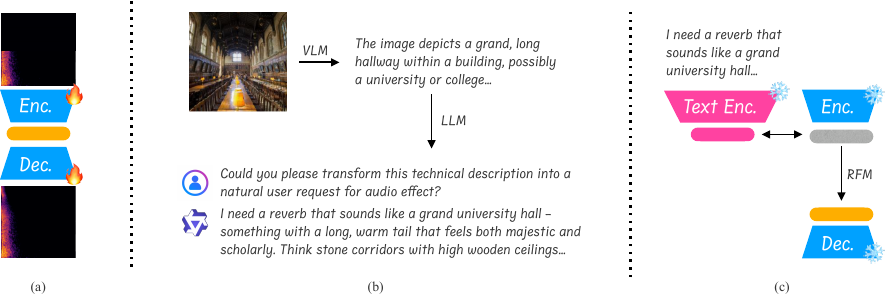}
    \caption{\textbf{\methodnamelong} system. 
    \textbf{(a)} A VAE pretraining stage that learns an encoder (Enc.) to produce compact latent representations and a decoder (Dec.) to upsample all impulse responses to 48 kHz;
    \textbf{(b)} A caption-then-rewrite pipeline that employs vision-language models (VLM) to generate descriptions of visual scenes, followed by large language models (LLM) that transform these descriptions into diverse textual prompts;
    \textbf{(c)} A latent rectified flow matching (RFM) model that generates reverb characteristics in the latent space, conditioned on the text descriptions produced in stage (b).
    }
    \label{fig:teaser}
\end{figure*}

\section{Method}
\label{sec:method}
We decompose \methodnamelong~into three components, demonstrated in Fig.~\ref{fig:teaser}: a VAE that learns and decodes a compact full‑band RIR latent (Sec.~\ref{sec:vae}), a caption‑then‑rewrite pipeline that yields natural‑language conditioning from images (Sec.~\ref{sec:caption}); and a conditional rectified‑flow Diffusion Transformer (DiT) that generates RIR latents from multimodal inputs using the frozen VAE decoder (Sec.~\ref{sec:flow}). 

\subsection{Variational Autoencoder}
\label{sec:vae}
We train a reverb variational autoencoder (VAE) that operates on mono impulse responses to learn compact latent representations. Our architecture employs ResBlocks~\citep{he2016resnet} as the encoder, processing complex-valued spectrograms projected onto 128 mel-scaled bands of the input impulse responses. For the decoder, we adopt a hybrid architecture combining 1D ConvNeXt blocks~\citep{liu2022convnext} with a preprocessing stage of Transformer~\citep{peebles2023dit,vaswani2017attention}, following the WavTokenizer~\citep{ji2024wavtokenizer}. The encoder employs a downsampling stride factor of $[1,2,3]$ on temporal axis, yielding a compact latent representation with dimensionality 16 and temporal resolution of 23.6 Hz. For a standard five-second impulse response, this configuration produces 118 latent frames. 

We train VAE following the $\beta$-VAE framework~\citep{higgins2017beta} with $\beta = 10^{-4}$, incorporating adversarial training using HiFi-GAN discriminators~\citep{kong2020hifigan}. Our total loss function comprises three components: a reconstruction loss computed on the Mel-spectrogram domain combined with the mean absolute error of RT60 values~\citep{singh2021image2reverb}, a Hinge GAN adversarial loss for improved perceptual quality, and a feature matching loss computed on intermediate discriminator feature maps. During training, we randomly reduce the sample rate for input audio, yet ask the decoder to still output full-band impulse response, thereby training it to be able to perform upsampling natively. This allows us to train the latent DiT model using only 22 kHz impulse response, yet still get full-band generation quality.

\subsection{Caption-Then-Rewrite Pipeline}
\label{sec:caption}
We begin by captioning images from~\citep{singh2021image2reverb} using two vision-language models (VLM): Moondream2 2B~\citep{moondream2024} and Qwen2-VL 72B Instruct~\citep{wang2024qwen2}. For Moondream2, we utilized its native captioning API directly. For Qwen2-VL, we designed a structured prompt requesting five acoustic-relevant fields: \emph{SPACE TYPE}, \emph{SIZE CLASS}, \emph{MAIN MATERIALS}, \emph{SOFT COVERAGE}, and \emph{RT60 BUCKET}, taking advantage of a large language backbone's extended reasoning ability and world knowledge.

To determine which description style better supports creative rewriting, we employed an LLM-as-a-judge evaluation with a 5-model panel: Gemini 2.5 Pro~\citep{comanici2025gemini}, GPT-4o~\citep{gpt4o2024}, GPT-5~\citep{gpt5024}, Claude Opus 4.1~\citep{claude2024}, and Mistral Medium 3.1~\citep{mistral2024}. Each judge performed pairwise comparisons between Qwen2.5-VL-72B~\citep{wang2024qwen2} and Moondream2~\citep{moondream2023} descriptions using the prompt: \textit{``Which image description is better for creative rewriting?''} Judges provided structured responses in \texttt{\textbackslash boxed\{A\}} or \texttt{\textbackslash boxed\{B\}} format, which we extracted via regular expressions. We evaluated 100 randomly sampled description pairs with randomized A/B positioning to prevent order bias. The panel reached near-unanimous consensus through majority voting, and to our surprise, the much smaller Moondream2 descriptions are preferred in 99.8\% of cases with exceptional inter-judge agreement. Despite this finding, we still opt to include Qwen2-VL's captioning results in, as we believe they provide key acoustic details for the space to guide generation.

Following VLM captioning, we employ Microsoft Phi-4 14B~\citep{phi42024} to transform factual descriptions into diverse natural user requests through multi-dimensional randomization. Our rewriting framework operates across several axes of variation to ensure comprehensive diversity in the generated prompts.

We incorporate 50+ distinct writing styles, spanning from \emph{``casual and conversational''} through \emph{``poetic and atmospheric''} to \emph{``technical and precise''}, ensuring broad linguistic coverage. We implement 40+ user personas representing diverse use cases, including \emph{``musician recording their first album''}, \emph{``experienced podcaster with millions of listeners''}, and \emph{``voice actor preparing for an animated film''}, capturing the heterogeneous nature of real-world RIR applications. Each generation follows structured templates that guide naturalistic transformation, such as \emph{``Transform this technical description into a natural user request for audio effects''} and \emph{``Convert this factual room description into how someone would ask for that acoustic sound''}. The system randomly selects conversational starters (\emph{``Hey,''} \emph{``I'd love,''} \emph{``Could you,''} \emph{``I'm hoping,''} \emph{``Please,''} \emph{``Can you help me''}) and appends emotional contexts (\emph{``- it's really important to me,''} \emph{``because it brings back memories,''} \emph{``to capture that perfect mood''}) to enhance authenticity.

For each image, we generate 55 creative prompts, including long and short prompt versions. We randomize temperature $\in [0.6, 1.2]$ and top-p $\in [0.8, 0.98]$ to balance creativity with coherence, ensuring linguistic diversity across generated prompts. The methodology successfully transforms technical specifications like \emph{``large concert hall; large; plaster, wood, upholstered-seats; 20; long (1.8 s)''} into authentic user requests such as \emph{``Hey, I need my audio to sound like it was recorded in a grand concert hall with that beautiful long reverb tail - it's really important to me,''} bridging the gap between acoustic parameters and natural language expression. 

\subsection{Rectified Flow Matching}
\label{sec:flow}

We train a DiT with conditional rectified flow matching~\citep{liu2022rectifiedflow} to learn the transformation from noise to RIR latents. The model learns a velocity field $v_\theta(\mathbf{x}_t,t,c)$ that transports Gaussian noise samples $\mathbf{x}_0\!\sim\!\mathcal{N}(0,I)$ to RIR latent representations $\mathbf{x}_1$ along optimal straight-line trajectories defined by $\mathbf{x}_t=(1-t)\mathbf{x}_0+t\mathbf{x}_1$ for $t\in[0,1]$. This transport follows the ordinary differential equation:
\begin{equation}
\frac{d\mathbf{x}}{dt}=v_\theta(\mathbf{x}_t,t,c),
\end{equation}
where $c$ represents multimodal conditioning including text embeddings, audio features, and constant parameters.

The training objective employs a flow-matching loss with pseudo-Huber penalty for improved gradient stability:
\begin{align}
\label{eq:fm}
\mathcal{L}_{\text{FM}} &=\mathbb{E}_{t,\mathbf{x}_0,\mathbf{x}_1,c}\!\left[L_\delta\!\big(v_\theta(\mathbf{x}_t,t,c)-(\mathbf{x}_1-\mathbf{x}_0)\big)\right], \\
L_\delta(\mathbf{z}) &=\delta^2\!\left(\sqrt{1+\|\mathbf{z}/\delta\|_2^2}-1\right), \nonumber
\end{align}
where we set $\delta=1.0$ per coordinate, matching the typical velocity scale in $\mathbf{x}_1-\mathbf{x}_0$. This parameterization remains invariant to data-parallel training strategies and GPU configurations.

To enable classifier-free guidance (CFG) during inference, we implement dropout-based conditioning augmentation during training, replacing $c$ with learned unconditional embeddings $c_{\text{uncond}}$ with probability $0.2$. At inference, we apply CFG with scale $6.0$ to enhance conditioning adherence:
\begin{align}
v_{\text{guided}} &= v_\theta(\mathbf{x}_t,t,c_{\text{uncond}}) \nonumber \\
&\quad + 6.0\,\big( v_\theta(\mathbf{x}_t,t,c) - v_\theta(\mathbf{x}_t,t,c_{\text{uncond}}) \big).
\end{align}

For generation, we integrate the learned ODE using an adaptive midpoint solver (RK2) with maximum 50 function evaluations and tolerance \texttt{rtol}$=\texttt{atol}=10^{-5}$. We apply cosine time reparameterization $t\mapsto\tau(t)=\tfrac{1-\cos(\pi t)}{2}$ during both training and sampling phases to improve convergence dynamics.

\section{Experiments}

\subsection{Dataset}
We assembled a large RIR dataset comprising 145,976 training, 7,964 validation, and 1,957 test samples, each standardized to 5-second duration. Our corpus integrates diverse acoustic environments from multiple established sources: C4DM~\citep{stewart2010database}, RIRS NOISES~\citep{ko2017study}, Image2Reverb~\citep{singh2021image2reverb}, synthetic RIRs generated via PyRoomAcoustics~\citep{scheibler2018pyroomacoustics}, SoundSpaces 2.0 RIRs~\citep{chen2022soundspaces}, and OpenAIR~\citep{murphy2010openair}. To enhance acoustic diversity beyond conventional spaces, we augmented the dataset with manually collected free-license RIRs from online forums, including unconventional environments such as ``RIR of inside a large glass fishbowl helmet\footnote{\href{https://creazilla.com/media/audio/15488033/dekker-fractanimal-impulse-response-large-glass-fishbowl-helmet}{Link to large glass fishbowl audio sample.}}.''

The complete training set of 145,976 samples was utilized to maximize acoustic diversity during model training. For multi-channel RIRs, we randomly select one channel at each training iteration.

\subsection{Objective Experiments}

\noindent \textbf{Text Encoder for Conditioning.} To identify optimal text encoding strategies for capturing acoustic semantics, we evaluated 15+ text encoders with varying pooling strategies. We investigated multiple pooling approaches: mean pooling (averaging across sequence dimension), maximum pooling (element-wise maximum across sequences), first-token extraction (utilizing the initial token representation), last-token pooling, and hybrid combinations. Our evaluation tested each configuration on 15 prompts spanning small intimate spaces to large concert halls. We assess encoder quality through three complementary metrics: (1) \textit{Batch Diversity}, measuring mean per-dimension standard deviation across the batch using raw embeddings; (2) \textit{Embedding Richness}, computing mean within-vector standard deviation across dimensions on raw embeddings; and (3) \textit{Semantic Separation}, calculating within-minus-between mean cosine similarity on L2-normalized embeddings using heuristic acoustic classes. Higher values indicate better performance for all metrics.

Table~\ref{tab:encoder_diversity} presents top-performing configurations from four architecture families: T5~\citep{raffel2020exploring}, BERT~\citep{devlin2018bert}, DeBERTa~\citep{he2020deberta}, and CLAP~\citep{elizalde2023clap}. Each architecture is shown with its best pooling strategy from preliminary experiments. Due to space constraints, we report only the highest-performing variant from each model family among 15+ total configurations evaluated. 

\begin{table*}[htbp]
\centering
\footnotesize
\begin{tabular}{lccccccccc}
\toprule
\textbf{Error Type} & \textbf{Baseline} & \textbf{XL, Long} & \textbf{XL, Short} & \textbf{L, Long} & \textbf{L, Short} & \textbf{B, Long} & \textbf{B, Short} & \textbf{S, Long} & \textbf{S, Short} \\
\midrule
Mean Error (\%) & -37.0 & 8.8 & 4.8 & 24.6 & 26.0 & 30.2 & 27.7 & 43.4 & 21.9 \\
Median Error (\%) & -59.6 & -33.95 & -38.5 & -19.0 & -23.9 & -16.7 & -18.6 & -10.4 & -23.6 \\
\bottomrule
\end{tabular}
\caption{RT60 estimation error analysis (n=1957). Negative values indicate underestimation.}
\label{tab:t60_errors}
\end{table*}

\vspace{-6pt}
\begin{table}[htbp]
\centering
\small
\begin{tabular}{@{}lccc@{}}
\toprule
\textbf{Configuration} & \textbf{Batch Div.} & \textbf{Embed. Rich.} & \textbf{Sem. Sep.} \\
\midrule
T5-Base + First & 0.228 & 0.281 & 0.079 \\
T5-Large + First & 0.163 & 0.219 & 0.095 \\
DeBERTa-v3 + Max & 0.279 & 1.682 & 0.003 \\
BERT-Base + Max & 0.235 & 0.390 & 0.010 \\
CLAP & 0.125 & 0.290 & 0.050 \\
\bottomrule
\end{tabular}
\caption{Text encoder evaluation for reverb-prompt embeddings across diversity and semantic separation metrics.}
\label{tab:encoder_diversity}
\end{table}

\vspace{6pt}

\noindent \textbf{VAE Reconstruction Quality.} Many impulse response generation work operate on spectrogram or mel-spectrogram space, then use Griffin-Lim to recover the phase information~\citep{steinmetz2018neuralreverberator}. As such, we evaluate our VAE against two Griffin-Lim baselines~\citep{griffin1984signal}: reconstruction from mel-spectrograms (GL-Mel) and from STFT magnitudes (GL-STFT). Table~\ref{tab:vae_vs_gl_core} presents the evaluation results. Our VAE demonstrates superior time-domain reconstruction fidelity with SNR of $-0.75$\,dB compared to $-5.26$–$-5.30$\,dB for Griffin-Lim variants, and MSE of $2.83\times10^{-4}$ versus $7.49$–$7.62\times10^{-4}$. Additionally, the VAE achieves approximately $62\times$ faster inference speed, enabling real-time applications. While the VAE exhibits higher RT60 error (mean $6.51\%$), this represents substantial improvement from initial training ($7.26\%$), indicating learned acoustic modeling capabilities beyond simple magnitude preservation~\citep{le2019blueprint, schroeder1965new}.
\vspace{-6pt}
\begin{table}[t]
\centering
\small
\setlength{\tabcolsep}{4pt}
\begin{tabular}{lcccc}
\toprule
 & SNR (dB) $\uparrow$ & MSE $\downarrow$ & RT60 (\%) $\downarrow$ & Time (ms) $\downarrow$ \\
\midrule
VAE (Ours) & -0.75 & 2.83 & 6.51 & 9.8 \\
GL--Mel    & -5.26 & 7.49 & 58.18 & 610.7 \\
GL--STFT   & -5.30 & 7.62 & 0.19  & 604.9 \\
\bottomrule
\end{tabular}
\caption{Comparison of Reconstruction Quality (means over $n{=}1957$). MSE is reported in units of $10^{-4}$.}
\label{tab:vae_vs_gl_core}
\end{table}

\vspace{6pt}

\noindent \textbf{Comparison with Baseline Methods.} Table~\ref{tab:rt60_results} analyzes RT60 predictions across the full test set (n=1957). Image2Reverb severely underestimates reverberation, achieving mean RT60 of $1.274$ seconds ($61.4\%$ deviation from ground truth $3.299$ seconds) with constrained maximum of $2.685$ seconds and reduced variance ($0.535$ seconds std). \methodnameshort~demonstrates superior acoustic modeling with clear scaling effects. The XL model achieves optimal performance: $2.189$ seconds for long prompts ($29.3\%$ deviation) and $2.044$ seconds for short prompts ($34.0\%$ deviation). Notably, larger models excel with long prompts while smaller models perform better with short prompts. All \methodnameshort~variants exhibit realistic dynamic range (maximum values $4.888$-$5.619$ seconds vs. ground truth $5.819$ seconds) and variance ($1.231$-$1.427$ seconds std vs. ground truth $1.715$ seconds), substantially outperforming Image2Reverb's constrained predictions and indicating faithful representation of diverse acoustic environments.

\vspace{-6pt}
\begin{table}[htbp]
\centering
\scalebox{1.0}{
\begin{tabular}{lccc}
\toprule
\textbf{Method} & \textbf{Mean (s)} & \textbf{Median (s)} & \textbf{Max (s)} \\
\midrule
Ground-Truth & $3.299$ & 3.106 & 5.819 \\
Image2Reverb~\citep{singh2021image2reverb} & $1.295$ & 1.211 & 2.685 \\
\midrule
\textit{\methodnameshort~}\\
XL, long & $2.189$ & 2.042 & 5.619 \\
XL, short & $2.044$ & 1.793 & 5.261 \\
L, long & $2.470$ & 2.454 & 5.472 \\
L, short & $2.378$ & 2.371 & 5.526 \\
B, long & $2.545$ & 2.431 & 5.455 \\
B, short & $2.472$ & 2.422 & 5.236 \\
S, long & $2.703$ & 2.681 & 5.516 \\
S, short & $2.391$ & 2.323 & 5.249 \\
\bottomrule
\end{tabular}
}
\caption{RT60 reverberation time vs. model size scaling (n=1957).}
\label{tab:rt60_results}
\end{table}

\vspace{6pt}
Following~\citep{singh2021image2reverb}, we compute RT60 errors as signed percentage deviations from ground truth. Table~\ref{tab:t60_errors} compares four \methodnameshort~model scales against Image2Reverb across the full test set, including S: 213M / 0.8GB / 1.9GB (12, 512, 8), B: 329M / 1.3GB / 2.4GB (16, 768, 12), L: 616M / 2.4GB / 3.6GB (24, 1024, 16), XL: 1.5B/ 5.9GB/ 7.7GB (32, 1536, 24), where the format shows params / RAM / model size and the triplet shows (\texttt{depth}, \texttt{hidden\_size}, \texttt{num\_heads}). Image2Reverb consistently underestimates RT60 with $\sim$$-37\%$ mean error.
\methodnameshort~shows strong scaling effects: S model ($43.4\%$ long, $21.9\%$ short), B model ($30.2\%$ long, $27.7\%$ short), L model ($24.6\%$ long, $26.0\%$ short), and XL model achieving breakthrough performance ($8.8\%$ long, $4.8\%$ short). The XL variant's near ground-truth accuracy across both prompt lengths significantly outperforms Image2Reverb and smaller models.

\subsection{Subjective Experiments}
To complement our quantitative analysis, we conducted a subjective evaluation with human listeners. Nine participants evaluated reverb quality and text-audio matching across three samples: ground truth recordings, Image2Reverb generations, and our \methodnameshort~method. Each participant rated 24 examples using a 5-point Likert scale, where 1 indicates "Very Poor" and 5 indicates "Excellent" quality or matching. Each trial presented a text description of the acoustic environment, followed by dry audio and three corresponding wet versions (reverberant). Participants rated each sample on two dimensions: overall reverb quality and how well the reverb matched expectations from the text.

\begin{figure}[htbp]
\centering
\begin{tikzpicture}
\begin{axis}[
    ybar,
    bar width=0.4cm,
    width=0.9\linewidth,  % Changed from 12cm
    height=0.55\linewidth,  % Proportional height
    ymin=0,
    ymax=5,
    ytick={1,2,3,4,5},
    yticklabels={Very Poor,Poor,Acceptable,Good,Excellent},
    xtick={1,2,3},
    xticklabels={Ground Truth,Image2Reverb,\methodnameshort~(Ours)},
    legend pos=north west,
    legend columns=2,
    grid=major,
    grid style={dashed,gray!30}
]
\addplot[fill=lightgray] coordinates {(1,3.32) (2,3.51) (3,3.79)};
\addplot[fill=lightblue] coordinates {(1,2.97) (2,3.26) (3,3.50)};
\legend{Reverb Quality, Text-Audio Match}
\end{axis}
\end{tikzpicture}
\caption{Subjective evaluation results showing mean ratings for reverb quality and text-audio matching. \methodnameshort~outperforms ground truth on both measures and Image2Reverb on quality.}
\label{fig:subjective_results}
\end{figure}
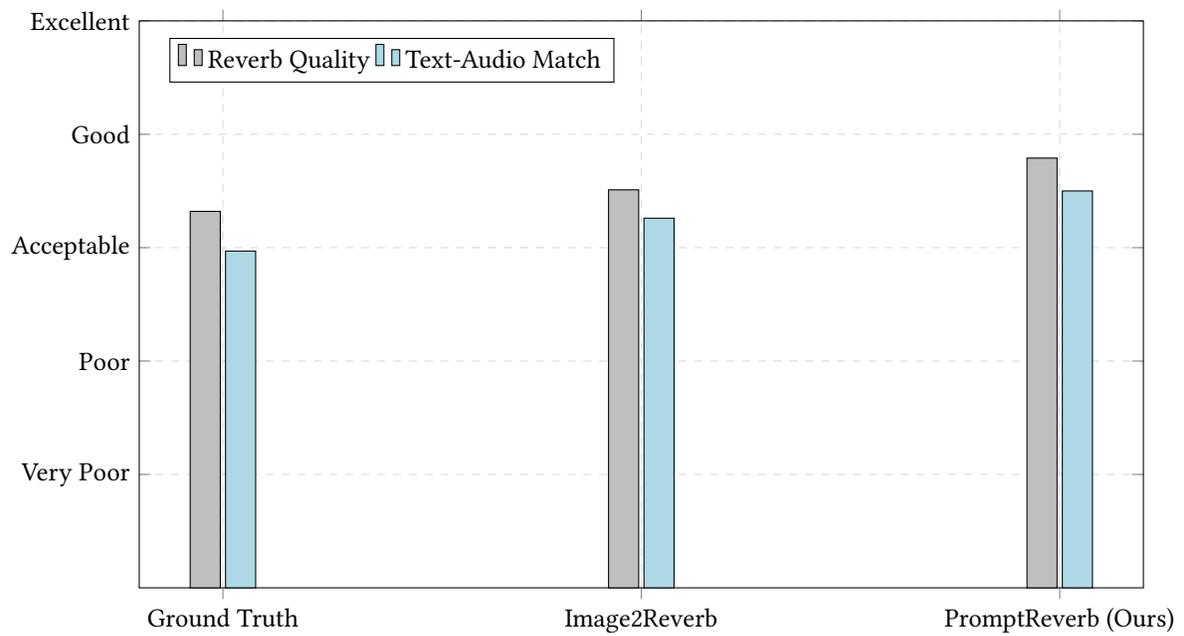

Statistical analysis using paired t-tests revealed \methodnameshort's advantages. For reverb quality, our method achieved 3.79 (±0.92) versus ground truth (3.32±1.24) and Image2Reverb (3.51±1.09). While statistically significant ($p<0.05$) against both, high standard deviations and small sample size make claimed superiority over real recordings questionable; however, the consistent improvement over Image2Reverb serves as a distant validation of the effectiveness of our approach. For text-audio matching, \methodnameshort~scored 3.50 (±0.98) compared to ground truth (2.97±1.21) and Image2Reverb (3.26±1.06). Despite statistical significance against ground truth ($p<0.05$), large variance warrants cautious interpretation.
% Comparable text-audio alignment with Image2Reverb indicates direct text conditioning maintains semantic fidelity while improving acoustic quality. 
Figure~\ref{fig:subjective_results} presents complete evaluation results.

% These results demonstrate that text-driven generation achieves comparable or better performance than image-based conditioning, with our method showing clear advantages over Image2Reverb in reverb quality. The comparable text-audio alignment between \methodnameshort~and Image2Reverb indicates that direct text conditioning maintains semantic fidelity while improving acoustic quality. Figure~\ref{fig:subjective_results} presents complete evaluation results.

\section{Conclusion}
\label{sec:conclusion}

We introduced \textit{\methodnamelong}, a two-stage framework that generates full-band RIRs from natural language descriptions. Our $\beta$-VAE achieves superior time-domain reconstruction; the rectified flow DiT generates acoustically accurate RIRs with 8.8\% mean RT60 error, outperforming Image2Reverb's $-37\%$ underestimation. Human evaluation confirms improved perceptual quality (3.79 vs. 3.51) while maintaining text-audio semantic alignment. \methodnameshort~represents the first system capable of synthesizing perceptually convincing 48 kHz RIRs from free-form text, eliminating requirements for panoramic imagery or technical acoustic expertise. This enables practical applications in VR/AR, game audio, and architectural acoustics where intuitive RIR generation is essential.

\clearpage
\newpage
\bibliographystyle{plainnat}
\bibliography{paper}

\end{document}